\documentclass[floats, prd, eqnum, showpacs, nofootinbib, 
twocolumn, 
eqsecnum]{revtex4-1}

\usepackage{color,graphicx}
\usepackage{amsfonts}
\usepackage{amssymb}
\usepackage{comment}
\begin{document}
    
\title{Deflection angle of a light ray reflected by a general marginally unstable photon sphere in a strong deflection limit}
\author{Naoki Tsukamoto${}^{1}$}\email{tsukamoto@rikkyo.ac.jp}
\affiliation{
${}^{1}$Department of General Science and Education, National Institute of Technology, Hachinohe College, Aomori 039-1192, Japan \\
}
\begin{abstract}
We investigate the deflection angle in a strong deflection limit for a marginally unstable photon sphere 
in a general asymptotically flat, static and spherically symmetric spacetime under some assumptions to calculate observables.
The deflection angle of a light ray reflected by the marginally unstable photon sphere diverges nonlogarithmically 
while the one reflected by a photon sphere diverges logarithmically.
We apply our formula to a Reissner-Nordstr\"{o}m spacetime and Hayward spacetime.
\end{abstract}

\maketitle

\section{Introduction}
Recently, the direct detection of 
gravitational waves emitted by black holes has been reported by
LIGO and VIRGO Collaborations \cite{Abbott:2016blz,LIGOScientific:2018mvr} and 
the detection of shadow of black hole candidate at center of a giant elliptical galaxy M87 
has been reported by Event Horizon Telescope Collaboration~\cite{Akiyama:2019cqa}.
From further observations,   
we will check more details of physics in strong gravitational fields in nature.

Static, spherically symmetric spacetimes, which describe strong gravitational fields caused by compact objects, have circular photon orbits and
the set of the photon orbits is called photon (antiphoton) sphere if it is unstable (stable)~\cite{Perlick_2004_Living_Rev}.
The property of the (anti)photon sphere such as the upper bound of the radius~\cite{Hod:2017xkz} and number~\cite{Hod:2017zpi} has been investigated 
since the (anti)photon sphere is related in phenomena in the strong gravitational field such as
the high-frequency behavior of the photon absorption cross section~\cite{Sanchez:1977si,Decanini:2010fz},
stability of thin-shell wormholes~\cite{Barcelo:2000ta,Koga:2020gqd}, 
the high-frequency spectrum of quasinormal modes of compact objects~\cite{Press:1971wr,Goebel_1972,Raffaelli:2014ola},
a centrifugal force and gyroscopic precession~\cite{Abramowicz_Prasanna_1990,Abramowicz:1990cb,Allen:1990ci,Hasse_Perlick_2002},
Bondi's sonic horizon of a radial fluid~\cite{Mach:2013gia,Chaverra:2015bya,Cvetic:2016bxi,Koga:2016jjq,Koga:2018ybs,Koga:2019teu},
and an apparent shape during a collapsing star to be a black hole~\cite{Ames_1968,Synge:1966okc,Yoshino:2019qsh}.
Extensions and alternatives of the photon sphere have been investigated~\cite{Claudel:2000yi,Koga:2019uqd,Cunha:2017eoe,Gibbons:2016isj,Shiromizu:2017ego,Yoshino:2017gqv,Galtsov:2019bty,Galtsov:2019fzq,Siino:2019vxh,Yoshino:2019dty,Cao:2019vlu,Yoshino:2019mqw,Lee:2020pre} 
and instability of the compact objects caused by the slow decay of linear waves near stable photon rings has been considered~\cite{Keir:2014oka,Cardoso:2014sna,Cunha:2017qtt}.

Gravitational lensing not only in weak gravitational fields~\cite{Schneider_Ehlers_Falco_1992,Schneider_Kochanek_Wambsganss_2006} 
but also in the strong gravitational field has been investigated eagerly.
Infinite number of dim images of light rays reflected by the photon sphere have been investigated many times~\cite{Hagihara_1931,Darwin_1959,Atkinson_1965,Luminet_1979,Ohanian_1987,Nemiroff_1993,
Frittelli_Kling_Newman_2000,Virbhadra_Ellis_2000,Bozza_Capozziello_Iovane_Scarpetta_2001,Bozza:2002zj,Perlick:2003vg,Nandi:2006ds,Virbhadra:2008ws,Bozza_2010,Tsukamoto:2016zdu,Shaikh:2019jfr,Wielgus:2020uqz}, 
which are named relativistic images in Ref.~\cite{Virbhadra_Ellis_2000},
to get the information of compact objects with the strong gravitational fields.

In 2002, Bozza has investigated a semianalytic formula to calculate the deflection angle of light rays reflected by a photon sphere, its image angle, and magnification 
in a general asymptotically flat, static and spherically symmetric spacetime~\cite{Bozza:2002zj}. 
The deflection angle of the light, in a strong deflection limit $b \rightarrow b_{\mathrm{m}}$, where $b$ and $b_{\mathrm{m}}$ are the impact parameter and the critical impact parameter of the light, respectively,
is given by
\begin{eqnarray}\label{eq:def0}
\alpha_{\mathrm{def}}&=&-\bar{a} \log \left( \frac{b}{b_{\mathrm{m}}}-1 \right) + \bar{b} \nonumber\\
&&+ O\left( \left( \frac{b}{b_{\mathrm{m}}}-1 \right)  \log \left( \frac{b}{b_{\mathrm{m}}}-1 \right) \right),
\end{eqnarray}
where $\bar{a}$ and $\bar{b}$ are parameters.
The formalism in the strong deflection limit has been extend by several authors~\cite{Bozza:2002af,Eiroa:2002mk,Petters:2002fa,Eiroa:2003jf,Bozza:2004kq,Bozza:2005tg,Bozza:2006sn,Bozza:2006nm,Iyer:2006cn,Bozza:2007gt,
Tsukamoto:2016qro,Ishihara:2016sfv,Tsukamoto:2016oca,Tsukamoto:2016zdu,Tsukamoto:2016jzh,Tsukamoto:2017edq,Shaikh:2018oul,Shaikh:2019itn}.
Relationships between the strong deflection limit, quasinormal modes~\cite{Stefanov:2010xz}, and high-energy absorption cross section~\cite{Wei:2011zw} have been investigated.

In this paper, we construct the deflection angle in the strong deflection limit for a marginally unstable photon sphere 
in a general asymptotically flat, static and spherically symmetric spacetime under some assumptions.
In the marginally unstable photon sphere case,  Eq.~(\ref{eq:def0}) is not suitable because the parameter $\bar{a}$ diverges.
We show that the deflection angle $\alpha_{\mathrm{def}}(b)$ is given by
\begin{equation}\label{eq:def01}
\alpha_{\mathrm{def}}(b)=\frac{\bar{c}}{\left(\frac{b}{b_{\mathrm{m}}}-1\right)^\frac{1}{6} } +\bar{d} +O\left( \left( \frac{b}{b_{\mathrm{m}}}-1\right)^\frac{1}{6} \right), 
\end{equation}
where $\bar{c}$ and $\bar{d}$ are constant.
We apply the formula to the Reissner-Nordstr\"{o}m spacetime and Hayward spacetime~\cite{Hayward:2005gi}.

Very recently, the deflection angle of a light reflected 
by a marginally unstable photon sphere at a wormhole throat in the Damour-Solodukhin wormhole spacetime~\cite{Damour:2007ap,Nandi:2018mzm,Ovgun:2018fnk,Bhattacharya:2018leh,Ovgun:2018swe} 
 in the strong deflection limit has been obtained as, in Ref.~\cite{Tsukamoto:2020uay}, 
\begin{equation}\label{eq:def02}
\alpha_{\mathrm{def}}(b)=\frac{\bar{c}_{[87]}}{\left(\frac{b}{b_{\mathrm{m}}}-1\right)^\frac{1}{4} } +\bar{d}_{[87]} +O\left( \left( \frac{b}{b_{\mathrm{m}}}-1\right)^\frac{3}{4} \right), 
\end{equation}
where $\bar{c}_{[87]}$ and $\bar{d}_{[87]}$ are constant, 
and the deflection angle of a light ray in a spacetime with naked singularity, which is suggested by Joshi~\textit{et al.} \cite{Joshi:2020tlq}, has been obtained as, in Ref.~\cite{Paul:2020ufc}, 
\begin{equation}\label{eq:def03}
\alpha_{\mathrm{def}}(b)=\frac{\bar{c}_{[89]}}{\left(\frac{b}{b_\mathrm{cr}}-1\right)^\frac{3}{2} } +\bar{d}_{[89]} +O\left( \left( \frac{b}{b_\mathrm{cr}}-1\right) \right),
\end{equation}
where $\bar{c}_{[89]}$, $\bar{d}_{[89]}$, and $b_\mathrm{cr}$ are constant.
We notice that results on this paper are compatible with ones in Refs.~\cite{Tsukamoto:2020uay,Paul:2020ufc} 
since the spacetimes are not satisfied our assumptions.
Note that we cannot apply our formula to light rays near a marginally unstable photon sphere if the marginally unstable photon sphere is correspond to a wormhole throat.

This paper is organized as follows. 
In Sec.~II, we investigate the deflection angle of a light in a general asymptotically flat, static and spherically symmetric spacetime with a marginally unstable photon sphere in the strong deflection limit under some assumptions.
In Sec.~III, we obtain a formula of observables in the strong deflection limit. 
We apply the formula to a Reissner-Nordstr\"{o}m spacetime and Hayward spacetime in Sec.~IV 
and we discuss and summarize our result in Sec.~V. 
On this paper we use the units in which a light speed and Newton's constant are unity.

\section{Deflection Angle in the strong deflection angle}
The line element of an asymptotically flat, static and spherically symmetric spacetime is given by
\begin{equation}
ds^2=-A(r)dt^2+B(r)dr^2+C(r)(d\vartheta^2+\sin^2 \vartheta d\varphi^2),
\end{equation}
where an asymptotically flat condition 
\begin{eqnarray}\label{eq:aym_con1}
&&\lim_{r\rightarrow \infty} A(r)=\lim_{r\rightarrow \infty} B(r)=1,\\\label{eq:aym_con2}
&&\lim_{r\rightarrow \infty} C(r)=r^2
\end{eqnarray}
is assumed. 
We assume an marginal unstable photon sphere at $r=r_\mathrm{m}$, which satisfies 
\begin{eqnarray}
&&D_\mathrm{m}=D^{\prime}_\mathrm{m}=0,\\
&&D^{\prime \prime}_\mathrm{m}=\frac{C^{\prime \prime \prime}_\mathrm{m}}{C_\mathrm{m}}-\frac{A^{\prime \prime \prime}_\mathrm{m}}{A_\mathrm{m}}>0,
\end{eqnarray}
where 
\begin{eqnarray}
D(r)\equiv \frac{C^{\prime}}{C}-\frac{A^{\prime}}{A},
\end{eqnarray}
and where $'$ denotes a differentiation with respect to $r$.
Here $D^{\prime}_\mathrm{m}$ is given by
\begin{eqnarray}
D^{\prime}_\mathrm{m}=\frac{C^{\prime \prime}_\mathrm{m}}{C_\mathrm{m}}-\frac{A^{\prime \prime}_\mathrm{m}}{A_\mathrm{m}}=0.
\end{eqnarray}
Here and hereafter, a function with the subtract $m$ denotes the function at $r=r_\mathrm{m}$.
We also assume that $r=r_\mathrm{m}$ is the largest positive solution of $D(r)=0$ and $A(r)$, $B(r)$, and $C(r)$ are positive and finite in a domain $r\geq r_\mathrm{m}$.
Time translational and axial Killing vectors $t^\mu \partial_\mu=\partial_t$ and $\varphi^\mu \partial_\mu=\partial_\varphi$ exist 
because of stationarity and axial symmetry, respectively.

By using the wave number of a photon $k^\mu \equiv \dot{x}^\mu$, where $x^\mu$ is the coordinates and where $\dot{\:}$ denotes a differentiation with respect to an affine parameter along a trajectory of a light ray,
the trajectory of the light ray is described by $k^\mu k_\mu=0$.
We assume $\vartheta=\pi/2$ without loss of generality because of spherical symmetry.
The trajectory of the light is rewritten as
\begin{equation}\label{eq:traj_1}
-A(r)\dot{t}^2+B(r)\dot{r}^2+C(r)\dot{\varphi}^2=0.
\end{equation}
By using an effective potential 
\begin{equation}
V(r)\equiv -\frac{L^2 F(r)}{B(r)C(r)},
\end{equation}
where $F(r)$ is defined by
\begin{equation}
F(r) \equiv \frac{C(r)}{A(r)b^2}-1,
\end{equation}
$b\equiv L/E$ is the impact parameter of the light, and where $E\equiv -g_{\mu \nu}t^{\mu}k^{\nu}$ and $L\equiv g_{\mu \nu} \varphi^\mu k^\nu$ are the conserved energy and angular momentum of the light, respectively, 
the trajectory of the light is expressed by
\begin{equation}
\dot{r}^2+V(r)=0.
\end{equation}
For simplicity, we concentrate on the positive impact parameter $b>0$.
\footnote{
Given the position of a point source, the lensed images with negative impact parameters can also be formed in the static, spherical symmetric spacetime. See Sec.~V.}
The light can be in a region with $V(r)\leq 0$.
From Eqs.~(\ref{eq:aym_con1}) and (\ref{eq:aym_con2}),
we obtain $\lim_{r\rightarrow \infty} V(r)=E^2>0$.
Thus, the light can be at a spatial infinity.
The first, second, and third derivatives of $V(r)$ with respective to $r$ are given by
\begin{eqnarray}
V^{\prime}&=&-\frac{L^2}{BC} \left( FG+\frac{CD}{Ab^2} \right), \\
V^{\prime \prime}&=&\frac{L^2}{BC} \left[ F\left(G^2-G^{\prime}\right)-\frac{C\left(D^2+D^{\prime}\right)}{Ab^2} \right],
\end{eqnarray}
and 
\begin{eqnarray}
V^{\prime \prime \prime}&=&\frac{L^2}{BC} \left[ F \left( -G^3+3GG^{\prime}-G^{\prime \prime} \right)  \right. \nonumber\\
&&+\frac{CD \left( -D^2+DG-3D^{\prime}+G^2-G^{\prime} \right)}{Ab^2} \nonumber\\
&&\left. +\frac{C\left( D^{\prime}G-D^{\prime \prime}\right)}{Ab^2} \right],
\end{eqnarray}
respectively,
where $G=G(r)$ is defined by
\begin{eqnarray}
G(r)\equiv \frac{B^{\prime}}{B}+\frac{C^{\prime}}{C}
\end{eqnarray}
and we have used a relation
\begin{eqnarray}
F^{\prime}=\frac{CD}{Ab^2}.
\end{eqnarray}

We assume that the light ray comes from the spatial infinity, it is reflected at a closest distance $r=r_0$ from a lensing object, and it goes to the spatial infinity.
Note that 
\begin{eqnarray}
&&\dot{r}_0 \equiv \left. \dot{r} \right|_{r=r_0}=0, \\
&&V_0 \equiv V(r_0)=0, \\
&&F_0 \equiv F(r_0)=0
\end{eqnarray}
at the closest distance $r=r_0$. 
Here and hereafter, a function with the subtract $0$ denotes the function at the closest distance $r=r_0$.
From Eq.~(\ref{eq:traj_1}), we obtain
\begin{equation}\label{eq:traj_cd}
A_0\dot{t}_0^2=C_0\dot{\varphi}^2_0.
\end{equation}
From Eq.~(\ref{eq:traj_cd}), the impact factor is expressed as 
\begin{equation}\label{eq:imp_2}
b(r_0)=\frac{L}{E}=\frac{C_0 \dot{\varphi}_0}{A_0 \dot{t}_0}= \sqrt{\frac{C_0}{A_0}}
\end{equation}
since it is constant along the trajectory of the light.
Recall that we have assumed that the impact parameter is positive.

We name a limit $r_0 \rightarrow r_\mathrm{m}$ or $b \rightarrow b_\mathrm{m}$ strong deflection limit,
where $b_\mathrm{m}$ is the critical impact parameter of the light ray defined as
\begin{equation}\label{eq:imp_c}
b_\mathrm{m} \equiv \sqrt{\frac{C_\mathrm{m}}{A_\mathrm{m}}}.
\end{equation}
In the strong deflection limit to the marginal unstable photon sphere $r_0 \rightarrow r_\mathrm{m}$ or $b \rightarrow b_\mathrm{m}$,
we obtain 
\begin{eqnarray}
&&\lim_{r_0 \rightarrow r_\mathrm{m}}V_0^{\prime}=0, \\
&&\lim_{r_0 \rightarrow r_\mathrm{m}}V_0^{\prime \prime}=0,\\
&&\lim_{r_0 \rightarrow r_\mathrm{m}}V_0^{\prime \prime \prime}=-\frac{L^2D^{\prime \prime}_\mathrm{m}}{B_\mathrm{m}C_\mathrm{m}}<0.
\end{eqnarray}

We can rewrite the trajectory equation as 
\begin{equation}
\left( \frac{dr}{d\varphi} \right)^2 = \frac{CF}{B}
\end{equation}
and we obtain the deflection angle $\alpha_{\mathrm{def}}(r_0)$ of the light as
\begin{equation}
\alpha_{\mathrm{def}}(r_0)\equiv I_0-\pi,
\end{equation}
where $I_0=I(r_0)$ is defined by
\begin{equation}
I_0\equiv 2 \int^\infty_{r_0} \frac{dr}{\sqrt{\frac{CF}{B}}}.
\end{equation}

We introduce a variable~\footnote{Bozza defined a variable $z_{[54]}$ as
\begin{equation}
z_{[54]}\equiv \frac{A-A_0}{1-A_0}
\end{equation}
in Ref.~\cite{Bozza:2002zj}.
See Refs.~\cite{Tsukamoto:2016qro,Tsukamoto:2016jzh} for the discussion on the variable.}
\begin{equation}
z\equiv 1-\frac{r_0}{r}
\end{equation}
and rewrite $I_0$ as
\begin{equation}
I_0=\int^1_0 f(z,r_0)dz,
\end{equation}
where $f(z,r_0)$ is defined as
\begin{equation}
f(z,r_0)\equiv \frac{2r_0}{\sqrt{H(z,r_0)}},
\end{equation}
and where $H(z,r_0)$ is defined by
\begin{equation}
H(z,r_0)\equiv \frac{CF}{B}(1-z)^4.
\end{equation}
By using the expansion of $F$ in the power of $z$ 
\begin{eqnarray}
F&=&D_0 r_0 z + \left[ \frac{r_0^2}{2} \left( \frac{C^{\prime \prime}_0}{C_0}-\frac{A^{\prime \prime}_0}{A_0} \right)+ \left(1- \frac{A^{\prime}_0 r_0}{A_0} \right) D_0 r_0 \right] z^2 \nonumber\\
&&+\left[ D_0 r_0 \left( 1-\frac{2r_0 A^{\prime}_0}{A_0} +\frac{r_0^2 A_0^{\prime 2}}{A_0^2} -\frac{r_0^2 A^{\prime \prime}_0}{2A_0} \right)  \right. \nonumber\\
&&+\left( \frac{C^{\prime \prime}_0}{C_0}-\frac{A^{\prime \prime}_0}{A_0} \right)r_0^2 \left( 1- \frac{r_0 A^{\prime}_0}{2A_0} \right)  \nonumber\\
&&\left.+ \frac{r_0^3}{6} \left( \frac{C^{\prime \prime \prime}_0}{C_0}-\frac{A^{\prime \prime \prime}_0}{A_0} \right) \right] z^3 +O\left( z^4 \right),
\end{eqnarray}
the expansion of $H(z,r_0)$ in the power of $z$ is obtained as
\begin{eqnarray}
H(z,r_0)=\sum^\infty_{n=1} c_n (r_0) z^n,
\end{eqnarray}
where $c_1(r_0)$, $c_2(r_0)$, and $c_3(r_0)$ are given by
\begin{eqnarray}
c_1(r_0)=\frac{C_0D_0r_0}{B_0},
\end{eqnarray}
\begin{eqnarray}
c_2(r_0)=\frac{C_0}{B_0}\left[ \frac{r_0^2}{2} \left( \frac{C^{\prime \prime}_0}{C_0}-\frac{A^{\prime \prime}_0}{A_0} \right)- \left(3 +\frac{A^{\prime}_0 r_0}{A_0} \right) D_0 r_0 \right], \nonumber\\
\end{eqnarray}
and 
\begin{eqnarray}
c_3(r_0)&=&\frac{C_0}{B_0}\left[ D_0 r_0 \left( 3+\frac{2r_0 A^{\prime}_0}{A_0} +\frac{r_0^2 A_0^{\prime 2}}{A_0^2} -\frac{r_0^2 A^{\prime \prime}_0}{2A_0} \right)  \right. \nonumber\\
&&-\left( \frac{C^{\prime \prime}_0}{C_0}-\frac{A^{\prime \prime}_0}{A_0} \right)r_0^2 \left( 1 +\frac{r_0 A^{\prime}_0}{2A_0} \right) \nonumber\\
&&\left. +\frac{r_0^3}{6} \left( \frac{C^{\prime \prime \prime}_0}{C_0}-\frac{A^{\prime \prime \prime}_0}{A_0} \right) \right], 
\end{eqnarray}
respectively.~\footnote{The expansions of a function $J(r)$ and its inverse $1/J(r)$ in the power of $z$ are shown in Appendix~A.}
In the strong deflection limit $r_0 \rightarrow r_{\mathrm{m}}$, $c_1(r_0)$, $c_2(r_0)$, and $c_3(r_0)$  become
\begin{eqnarray}
c_{\mathrm{1m}}=c_{\mathrm{2m}}=0
\end{eqnarray}
and 
\begin{eqnarray}
c_{\mathrm{3m}}=\frac{C_{\mathrm{m}}}{6b_{\mathrm{m}}} D^{\prime \prime}_{\mathrm{m}} r_{\mathrm{m}}^3
\end{eqnarray}
and $H(z,r_0)$ becomes
\begin{eqnarray}
H(z,r_{\mathrm{m}})=c_{\mathrm{3m}}z^3 +O\left( z^4 \right).
\end{eqnarray}
Therefore, $f(z,r_0)$ diverges with the leading order of $z^{-\frac{3}{2}}$ and $I_0$ diverges with $z^{-\frac{1}{2}}$ in the strong deflection limit $r_0 \rightarrow r_{\mathrm{m}}$.

We separate $I_0$ into a divergence part 
\begin{eqnarray}
I_{\mathrm{D}}&\equiv& \int^1_0 f_{\mathrm{D}}(z,r_{\mathrm{m}})dz \nonumber\\
&=&\frac{4r_{\mathrm{m}}}{\sqrt{c_{\mathrm{3m}}}} \frac{1}{  \sqrt{\frac{r_0}{r_{\mathrm{m}}}-1 }}-\frac{4r_{\mathrm{m}}}{\sqrt{c_{\mathrm{3m}}}} +O\left( \sqrt{\frac{r_0}{r_{\mathrm{m}}}-1} \right), \nonumber\\
\end{eqnarray}
where $f_{\mathrm{D}}(z,r_{\mathrm{m}})$ is defined as
\begin{eqnarray}
f_{\mathrm{D}}(z,r_{\mathrm{m}})&\equiv& \frac{2r_{\mathrm{m}}}{\sqrt{c_{\mathrm{1m}} z+c_{\mathrm{2m}} z^2+c_{\mathrm{3m}} z^3}} \nonumber\\
&=&\frac{2r_{\mathrm{m}}}{\sqrt{c_{\mathrm{3m}} z^3}},
\end{eqnarray}
and a regular part 
\begin{eqnarray}
I_{\mathrm{R}}(r_0)&\equiv& \int^1_0 f_{\mathrm{R}}(z,r_0) dz,
\end{eqnarray}
where $f_{\mathrm{R}}$ is defined as
\begin{eqnarray}
f_{\mathrm{R}}(z,r_0)\equiv f(z,r_0)-f_{\mathrm{D}}(z,r_{\mathrm{m}}).
\end{eqnarray}
The regular part $I_{\mathrm{R}}$ is expanded in the power of $r_0-r_{\mathrm{m}}$ as
\begin{equation}
I_{\mathrm{R}}(r_0)=\sum^\infty_{j=0} \frac{1}{j!} \left( r_0-r_{\mathrm{m}} \right)^j \int^1_0 \left. \frac{\partial^j f_{\mathrm{R}}}{\partial r_0^j} \right|_{r_0=r_{\mathrm{m}}} dz 
\end{equation}
and we are interested in the term of $j=0$ given by
\begin{equation}
I_{\mathrm{R}}(r_0)= \int^1_0 f_{\mathrm{R}}(z,r_{\mathrm{m}}) dz +O\left( \sqrt{\frac{r_0}{r_{\mathrm{m}}-1}} \right). 
\end{equation}
By using an equation
\begin{equation}
b=b_{\mathrm{m}}+ \frac{b_{\mathrm{m}} D^{\prime \prime}_{\mathrm{m}}}{12} \left( r_0 -r_{\mathrm{m}} \right)^3 + O\left( \left( r_0 -r_{\mathrm{m}} \right)^4 \right),
\end{equation}
$I_{\mathrm{D}}$ is rewritten as
\begin{eqnarray}
I_{\mathrm{D}}(b)&=&\frac{2^\frac{13}{6} 3^\frac{1}{3} b_{\mathrm{m}}^\frac{1}{2}}{ C_{\mathrm{m}}^\frac{1}{2} D_{\mathrm{m}}^{\prime \prime \frac{1}{3}} \left( \frac{b}{b_{\mathrm{m}}}-1\right)^\frac{1}{6} } 
-4\sqrt{\frac{6 b_{\mathrm{m}}}{C_{\mathrm{m}} D^{\prime \prime}_{\mathrm{m}} r_{\mathrm{m}}}}\nonumber\\
&& +O\left( \left( \frac{b}{b_{\mathrm{m}}}-1 \right)^\frac{1}{6} \right). 
\end{eqnarray}
Therefore, we obtain the deflection angle $\alpha_{\mathrm{def}}(b)$ as
\begin{equation}
\alpha_{\mathrm{def}}(b)=\frac{\bar{c}}{\left(\frac{b}{b_{\mathrm{m}}}-1\right)^\frac{1}{6} } +\bar{d} +O\left( \left(\frac{b}{b_{\mathrm{m}}}-1\right)^\frac{1}{6} \right), 
\end{equation}
where 
\begin{eqnarray}
&&\bar{c}=\frac{2^\frac{13}{6} 3^\frac{1}{3} b_{\mathrm{m}}^\frac{1}{2}}{C_{\mathrm{m}}^\frac{1}{2} D_{\mathrm{m}}^{\prime \prime \frac{1}{3}}}, \\
&&\bar{d}=-4\sqrt{\frac{6b_{\mathrm{m}}}{C_{\mathrm{m}} D^{\prime \prime}_{\mathrm{m}} r_{\mathrm{m}}}} +I_{\mathrm{R}} -\pi. 
\end{eqnarray}

\section{Observables in the strong deflection limit}
We consider that 
a light ray with an impact parameter $b$ 
is emitted by a source S with a source angle $\phi$ 
and that it is reflected by a lens L with an effective deflection angle 
\begin{eqnarray}
\bar{\alpha}_{\mathrm{def}}\equiv \alpha_{\mathrm{def}} \quad \mathrm{mod} \;  2 \pi
\end{eqnarray}
to reach into an observer O.
A relation between a deflection angle $\alpha_{\mathrm{def}}$ and the effective deflection angle $\bar{\alpha}_{\mathrm{def}}$ can be expressed by
\begin{eqnarray}\label{eq:baralpha}
\alpha_{\mathrm{def}}=\bar{\alpha}_{\mathrm{def}}+ 2 \pi n,
\end{eqnarray}
where $n$ is the winding number of the light ray. 
The observer O observes an image I with an image angle $\theta$. 
See Figure 1 for a lens configuration.
\begin{figure}[htbp]
\begin{center}
\includegraphics[width=70mm]{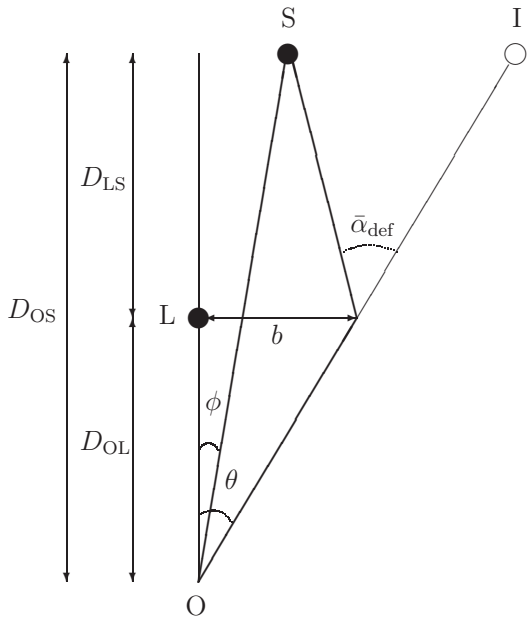}
\end{center}
\caption{Lens configuration. 
A light with an impact parameter $b$ is emitted by a source S with a source angle $\phi$ and it is reflected by a lens L with an effective deflection angle $\bar{\alpha}_{\mathrm{def}}$ to reach into an observer O.
The observer O sees image I with an image angle $\theta$. 
$D_{\mathrm{OS}}$, $D_{\mathrm{LS}}$, and $D_{\mathrm{OL}}$ are distances between the observer O and the source S, between the lens L and the source S, and between the observer O and the lens L, respectively.}
\end{figure}
We assume a small angle lens equation \cite{Bozza:2008ev}
\begin{eqnarray}\label{eq:lens}
D_{\mathrm{LS}}\bar{\alpha}_{\mathrm{def}}=D_{\mathrm{OS}}(\theta -\phi),
\end{eqnarray}
where $D_{\mathrm{LS}}$ and $D_{\mathrm{OS}}$ are angular distances between the lens L and the source S and between the observer O and the source S, respectively. 
We have assumed that all the angles $\bar{\alpha}_{\mathrm{def}}, \theta, \phi \ll 1$.
The impact parameter $b$ is described by 
\begin{eqnarray}
b=\theta D_{\mathrm{OL}},
\end{eqnarray}
where $D_{\mathrm{OL}}=D_{\mathrm{OS}}-D_{\mathrm{LS}}$ is an angular distance between the observer O and the lens L.
The deflection angle is rewritten as
\begin{equation}\label{eq:alpha}
\alpha_{\mathrm{def}}(\theta)=\frac{\bar{c}}{\left( \frac{\theta}{\theta_\infty}-1 \right)^\frac{1}{6} } +\bar{d} +O\left( \left( \frac{\theta}{\theta_\infty}-1 \right)^\frac{1}{6} \right), 
\end{equation}
where $\theta_\infty \equiv b_{\mathrm{m}}/D_{\mathrm{OL}}$.

We define $\theta^0_n$ as, for $n\geq 1$,
\begin{eqnarray}\label{eq:theta0n}
\alpha_{\mathrm{def}} \left(\theta^0_n \right)=2\pi n.
\end{eqnarray}
From Eqs. (\ref{eq:alpha}) and (\ref{eq:theta0n}), we get
\begin{eqnarray}\label{eq:theta0n2}
\theta^0_n = \left[ 1 + \left( \frac{\bar{c}}{2\pi n - \bar{d}} \right)^6 \right] \theta_\infty.
\end{eqnarray}
The deflection angle $\alpha_{\mathrm{def}}\left(\theta \right)$ can be expanded around $\theta=\theta^0_n$ as 
\begin{eqnarray}\label{eq:alphaexpand}
\alpha_{\mathrm{def}}(\theta) 
&=&\alpha_{\mathrm{def}} \left(\theta^0_n \right)+\left.\frac{d\alpha_{\mathrm{def}}}{d\theta}\right|_{\theta=\theta^0_n} \left(\theta-\theta^0_n \right) \nonumber\\
&&+O\left( \left( \theta-\theta^0_n \right)^2 \right), 
\end{eqnarray}
where, from Eq. (\ref{eq:alpha}), 
\begin{eqnarray}\label{eq:dalphadtheta}
\left.\frac{d\alpha_{\mathrm{def}}}{d\theta}\right|_{\theta=\theta^0_n}=-\frac{\bar{c}}{6\theta_\infty} \left( \frac{\theta^0_n}{\theta_\infty}-1 \right)^{-\frac{7}{6}}.
\end{eqnarray}
From Eqs. (\ref{eq:baralpha}), (\ref{eq:theta0n}), (\ref{eq:alphaexpand}), and (\ref{eq:dalphadtheta}), the effective deflection angle is given by
\begin{eqnarray}\label{eq:baralpha2}
\bar{\alpha}_{\mathrm{def}}\left(\theta_n \right)=\frac{\bar{c}}{6\theta_\infty} \left( \frac{\theta^0_n}{\theta_\infty} -1 \right)^{-\frac{7}{6}} \left( \theta^0_n - \theta_n \right),
\end{eqnarray}
where $\theta_n$ is an image angle with the winding number $n$.

From the lens equation (\ref{eq:lens}) and the effective deflection angle (\ref{eq:baralpha2}),
the image angle $\theta_n=\theta_n(\phi)$ with the winding number $n$ is obtained as 
\begin{eqnarray}
\theta_n(\phi)\sim \theta^0_n +\frac{6 \bar{c}^6 D_{\mathrm{OS}} \theta_\infty (\phi - \theta^0_n )}{D_{\mathrm{LS}} \left( 2 \pi n - \bar{d} \right)^7}
\end{eqnarray}
and an Einstein ring angle with the winding number $n$ is given by
\begin{eqnarray}
\theta_{\mathrm{E}n}\equiv \theta_n(0)\sim \theta^0_n \left[1-\frac{6 \bar{c}^6 D_{\mathrm{OS}} \theta_\infty }{D_{\mathrm{LS}} \left( 2 \pi n - \bar{d} \right)^7} \right].
\end{eqnarray}
The magnification of the image with $\theta_n$ is obtained by
\begin{eqnarray}
\mu_n(\phi) \equiv \frac{\theta_n}{\phi}\frac{d\theta_n}{d\phi}
 \sim \frac{6\bar{c}^6 D_{\mathrm{OS}} \theta_\infty^2}{D_{\mathrm{LS}} \phi} K_n,
\end{eqnarray}
where 
\begin{eqnarray}
K_n\equiv \frac{(2\pi n-\bar{d})^6+\bar{c}^6}{(2\pi n-\bar{d})^{13}}.
\end{eqnarray}
The ratio of the magnification of the outermost image with $n=1$ to the sum of the other images with $n\geq 2$ is given by 
\begin{eqnarray}
\bar{r}\equiv \frac{\mu_1}{\sum^\infty_{n=2} \mu_n} \sim \frac{K_1}{\sum^\infty_{n=2} K_n}
\end{eqnarray}
and the difference of the image angles between the outermost image and the marginally unstable photon sphere is given by
\begin{eqnarray}
\bar{s}\equiv \theta_1 - \theta_\infty \sim \theta^0_1 - \theta^0_\infty = \left( \frac{\bar{c}}{2\pi - \bar{d}} \right)^6 \theta_\infty.
\end{eqnarray}

\section{Application}
We apply the formula in Sections II and III to the Reissner-Nordstr\"{o}m spacetime and the Hayward spacetime 
with the marginally unstable photon spheres.

\subsection{Reissner-Nordstr\"{o}m spacetime}
Eiroa, Romero, and Torres~\cite{Eiroa:2002mk} have investigated deflection angle of a light in the Reissner-Nordstr\"{o}m spacetime in the strong deflection limit
by using a numerical fitting method to calculate it. They have considered the magnifications of images of not only a point source but also an extended source. 
In Ref.~\cite{Bozza:2002zj}, Bozza has investigated a formula in the strong deflection limit in a general static, spherically symmetric spacetime 
to apply it to the Reissner-Nordstrom black hole spacetime.
The time delay of light rays~\cite{Sereno:2003nd} and retrolensing in the Reissner-Nordstrom spacetime~\cite{Eiroa:2003jf,Tsukamoto:2016oca,Tsukamoto:2016jzh} have been investigated. 
In the overcharged cases, the size of the photon sphere or the shadow and its magnifications has been studied in Refs.~\cite{Zakharov:2014lqa,Shaikh:2019itn,Wielgus:2020uqz}.
The shadow of a reflection-asymmetric thin-shell wormhole, which is consists of Reissner-Nordstr\"{o}m geometries, also has been considered in Ref.~\cite{Wielgus:2020uqz}. 

The functions $A(r)$, $B(r)$, and $C(r)$ in the Reissner-Nordstr\"{o}m spacetime with a marginally unstable photon sphere are given by
\begin{eqnarray}
&&A(r)=\frac{1}{B(r)}=1-\frac{2M}{r}+\frac{9M^2}{8r^{2}},\\
&&C(r)=r^{2}.
\end{eqnarray}
There is the marginally unstable photon sphere at $r=r_{\mathrm{m}}$, 
where $r_{\mathrm{m}}$ is given by 
\begin{equation}
r_{\mathrm{m}}=\frac{3M}{2}
\end{equation}
and the critical impact parameter is given by
\begin{equation}
b_{\mathrm{m}}=\frac{3\sqrt{6}M}{2}.
\end{equation}
Figure~2 shows the effective potential $V/E^2$ as a function of $r/r_{\mathrm{m}}$.
\begin{figure}[htbp]
\begin{center}
\includegraphics[width=87mm]{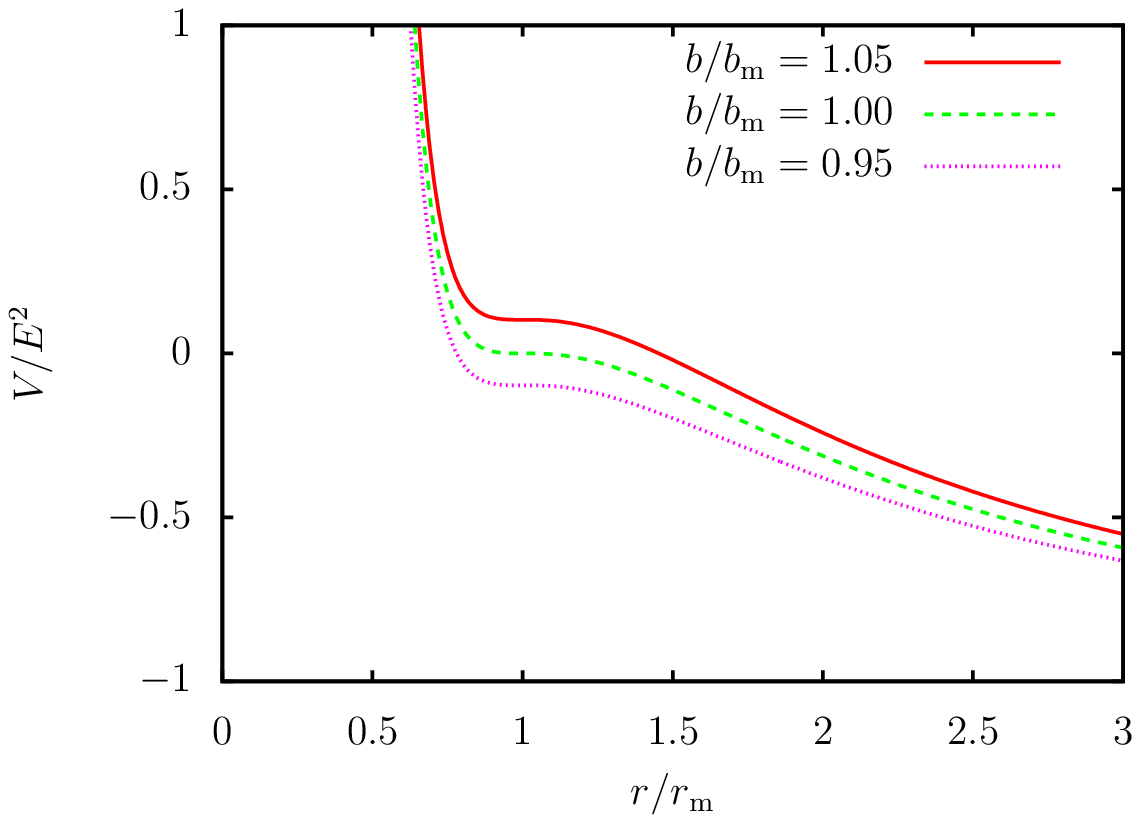}
\end{center}
\caption{The effective potential as a function of $r/r_{\mathrm{m}}$ in the Reissner-Nordstr\"{o}m spacetime.
The solid~(red), broken~(green), and dotted~(magenta) curves denote $V/E^2$ for $b/b_{\mathrm{m}}=1.05$, $1$, and $0.95$, respectively.}
\end{figure}

In the Reissner-Nordstr\"{o}m spacetime, 
$f_{\mathrm{R}}(z,r_{\mathrm{m}})$ has a simple form given by
\begin{eqnarray}
f_{\mathrm{R}}(z,r_{\mathrm{m}}) = 2\sqrt{\frac{6}{z^3(4-3z)}}-\sqrt{\frac{6}{z^3}}  
\end{eqnarray}
and we can obtain $I_{\mathrm{R}}(r_{\mathrm{m}})=\sqrt{6}$ analytically. 
Therefore, we obtain $\bar{c}$ and $\bar{d}$ analytically as
\begin{equation}
\bar{c}=2^\frac{5}{3}3^\frac{1}{2} \sim 5.49892
\end{equation}
and
\begin{equation}
\bar{d}= -\sqrt{6} -\pi \sim -5.59108,
\end{equation}
respectively.
Figure~3 shows the deflection angle $\alpha_{\mathrm{def}}(b/b_{\mathrm{m}})$ in the strong deflection limit~(2.50) and the one calculated by Eq.~(2.27) numerically. 
\begin{figure}[htbp]
\begin{center}
\includegraphics[width=87mm]{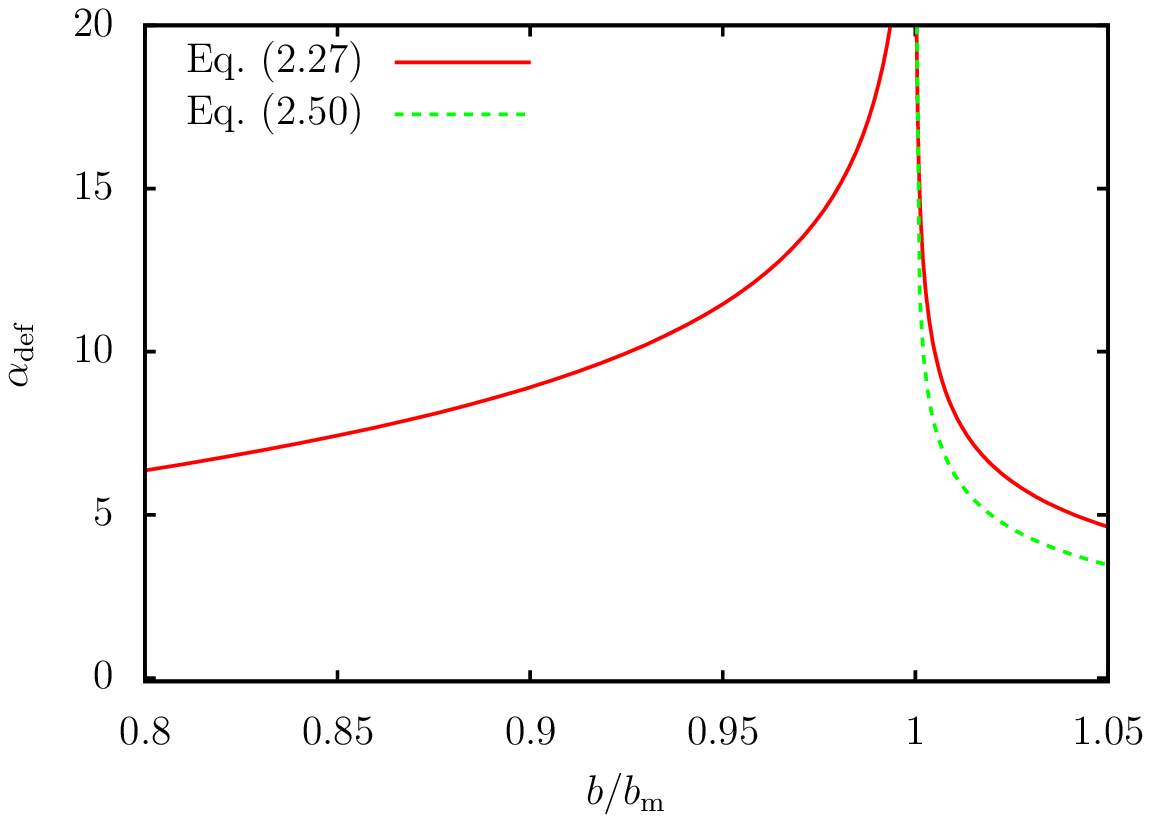}
\end{center}
\caption{The deflection angle as a function of $b/b_{\mathrm{m}}$ in the Reissner-Nordstr\"{o}m spacetime.
The solid~(red) and broken~(green) curves denote the deflection angle $\alpha_{\mathrm{def}}$ of Eqs.~(2.27) and (2.50), respectively.}
\end{figure}
The observables in the strong deflection limit are summarized in Table~I.
\begin{table*}[htbp]
 \label{table:I}
 \caption{Parameters $\bar{c}$ and $\bar{d}$ in the deflection angle in the strong deflection limit in the Reissner-Nordstr\"{o}m and Hayward spacetimes with the marginally unstable photon sphere, 
 the diameters of the relativistic Einstein rings $2\theta_{\mathrm{E}1}$, $2\theta_{\mathrm{E}2}$, and $2\theta_{\mathrm{E}3}$ with the winding number $n=1$, $2$, and $3$, respectively, and 
 the innermost ring~$2\theta_{\infty}$ scattered by the marginally unstable photon sphere, 
 the difference of the radii of the outermost ring and the innermost ring $\bar{s}=\theta_{\mathrm{E}1}-\theta_\infty$, 
 the total magnifications of the pair images $\mu_{1\mathrm{tot}} \sim 2 \left| \mu_{1} \right|$, 
 $\mu_{2\mathrm{tot}} \sim 2 \left| \mu_{2} \right|$, and $\mu_{3\mathrm{tot}} \sim 2 \left| \mu_{3} \right|$ with $n=1$, $2$, and $3$, respectively, 
 the ratio of the magnification of the outermost ring to the other rings $\bar{r}= \mu_1/\sum^\infty_{n=2} \mu_n$ in the case with distances
 $D_{\mathrm{OS}}=16$kpc and $D_{\mathrm{OL}}=D_{\mathrm{LS}}=8$kpc and with mass $M=4\times 10^6 M_{\odot}$, 
 and with the source angle $\phi=1$ arcsecond for $\mu_{1\mathrm{tot}}$, $\mu_{2\mathrm{tot}}$, and $\mu_{3\mathrm{tot}}$.}
\begin{center}
\begin{tabular}{c c c} \hline
                             &Reissner-Nordstr\"{o}m &Hayward   \\ \hline
 $\bar{c}$                   &5.49892                &4.95196 \\ 
 $\bar{d}$                   &-5.59108               &-5.62607 \\ 
 $2\theta_{\mathrm{E}1}$ [$\mu$as]     &36.8325                &46.4819 \\ 
 $2\theta_{\mathrm{E}2}$ [$\mu$as]     &36.5009                &46.2617 \\ 
 $2\theta_{\mathrm{E}3}$ [$\mu$as]     &36.4775                &46.2461 \\ 
 $2\theta_{\infty}$ [$\mu$as] &36.4727                &46.2429 \\
 $\bar{s}$ [$\mu$as]        &0.180                  &0.120 \\ 
 $\mu_{1\mathrm{tot}}\times 10^{19}$  &324.594                &271.346 \\ 
 $\mu_{2\mathrm{tot}}\times 10^{19}$  &16.4542                &13.9129 \\ 
 $\mu_{3\mathrm{tot}}\times 10^{19}$  &2.05397                &1.74330 \\ 
 $\bar{r}$		     &16.9947                &16.7916 \\ \hline
\end{tabular}
\end{center}
\end{table*}

\subsection{Hayward spacetime}
In Ref.~\cite{Wei:2015qca}, the deflection angle of a light in a weak-field approximation and in the strong deflection limit in the Hayward black hole spacetime~\cite{Hayward:2005gi} was considered. 
Chiba and Kimura~\cite{Chiba:2017nml} showed the shadow images of in the black hole and its over-charged case.

The functions
$A(r)$, $B(r)$, and $C(r)$ in the Hayward spacetime with a marginally unstable photon sphere are given by
\begin{eqnarray}
&&A(r)=\frac{1}{B(r)}=1-\frac{2Mr^2}{r^3+2q^2M},\\
&&C(r)=r^{2},
\end{eqnarray}
where $q=\pm \frac{25\sqrt{30}}{144}M$.
The marginal unstable photon sphere is at $r=r_{\mathrm{m}}=25M/12$ 
and the critical impact parameter $b_{\mathrm{m}}$ is obtained as
\begin{equation}
b_{\mathrm{m}}=\frac{25\sqrt{5}M}{12}.
\end{equation}
The effective potential $V/E^2$ as a function of $r/r_{\mathrm{m}}$ is shown as Fig.~4.
\begin{figure}[htbp]
\begin{center}
\includegraphics[width=87mm]{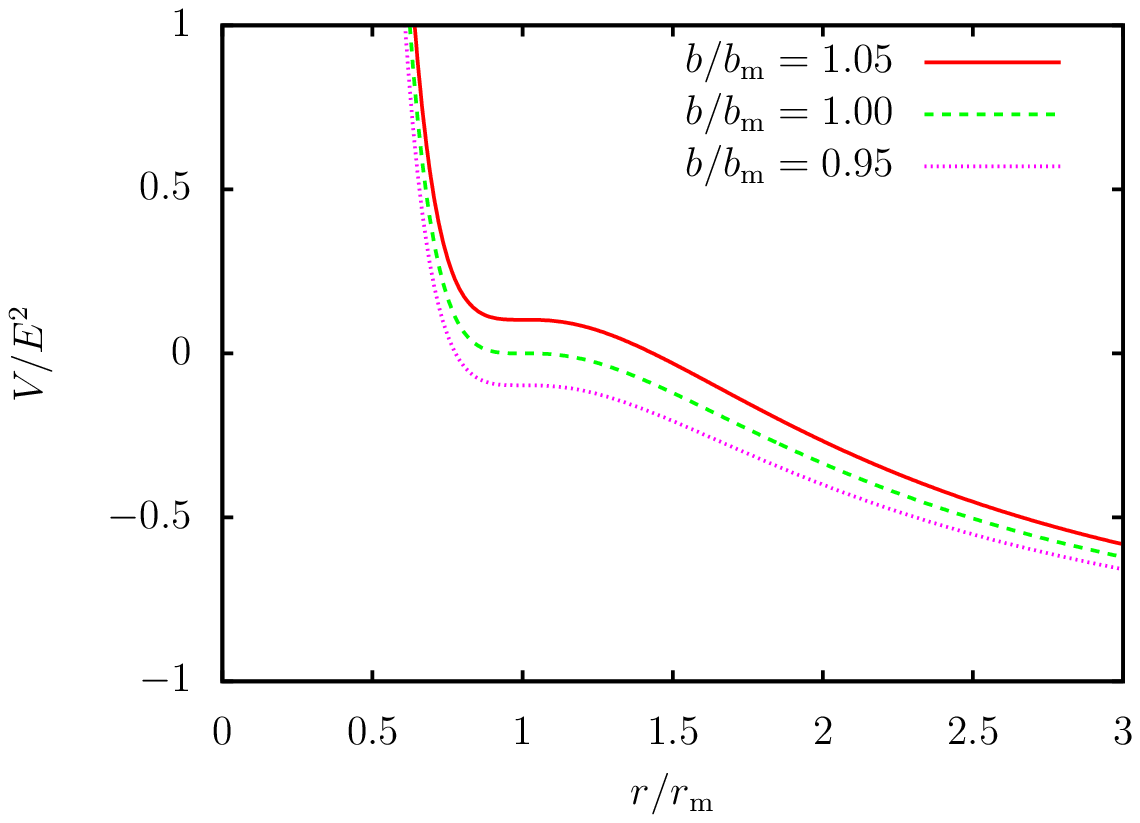}
\end{center}
\caption{The effective potential as a function of $r/r_{\mathrm{m}}$ in the Hayward spacetime.
The solid~(red), broken~(green), and dotted~(magenta) curves denote $V/E^2$ for $b/b_{\mathrm{m}}=1.05$, $1$, and $0.95$, respectively.}
\end{figure}

In the Hayward spacetime, $f_{\mathrm{R}}(z,r_{\mathrm{m}})$ is given by
\begin{eqnarray}
f_{\mathrm{R}}(z,r_{\mathrm{m}}) = 2\sqrt{{\frac{6-3z+3z^2-z^3}{z^3(5-5z+z^2)}}} -2\sqrt{\frac{6}{5z^3}}.
\end{eqnarray}
It is integrated numerically and it gives $I_{\mathrm{R}}(r_{\mathrm{m}})=1.8973$. 
By applying the formula in Sec.~II to the Hayward spacetime, we obtain
\begin{eqnarray}
\bar{c}=\frac{2^\frac{13}{6} 3^\frac{1}{3}}{5^\frac{1}{6}} \sim 4.95196
\end{eqnarray}
and
\begin{equation}
\bar{d}= -4\sqrt{\frac{6}{5}}+I_{\mathrm{R}}(r_{\mathrm{m}}) -\pi \sim -5.62607.
\end{equation}

Chiba and Kimura~\cite{Chiba:2017nml} obtained the deflection angle $\alpha_{\mathrm{def}}(b)$ as
\begin{equation} 
\alpha_{\mathrm{def}}(b)\sim \frac{c_{[94]} M^\frac{1}{6}}{b_{\mathrm{m}}^{\frac{1}{6}}  \left( \frac{b}{b_{\mathrm{m}}}-1 \right)^\frac{1}{6} } \sim \frac{6.01316}{\left( \frac{b}{b_{\mathrm{m}}}-1 \right)^\frac{1}{6} }, 
\end{equation}
where 
\begin{equation}
c_{[94]}=2^{\frac{11}{6}}3^{\frac{2}{3}}5^{\frac{5}{4}} \int^\infty_0\frac{dy}{\sqrt{432y+900y^2+625y^3}} \sim 7.771.
\end{equation}
See Eqs.~(21) and (22) in Ref.~\cite{Chiba:2017nml}.
Our formula~(2.50) recovers the form of Eq.~(4.14).
Figure~5 shows the deflection angle~$\alpha_{\mathrm{def}}(b/b_{\mathrm{m}})$ in the strong deflection limit (2.50), the deflection angle (2.27) calculated in numerical, and the deflection angle~(4.14) shown in Ref.~\cite{Chiba:2017nml}.
Equation~(4.14) is well matched with the numerical calculation of Eq.~(2.27) only near $b=b_{\mathrm{m}}$ since it has a divergent term only.
The observables in the strong deflection limit are shown in Table~I.
\begin{figure}[htbp]
\begin{center}
\includegraphics[width=87mm]{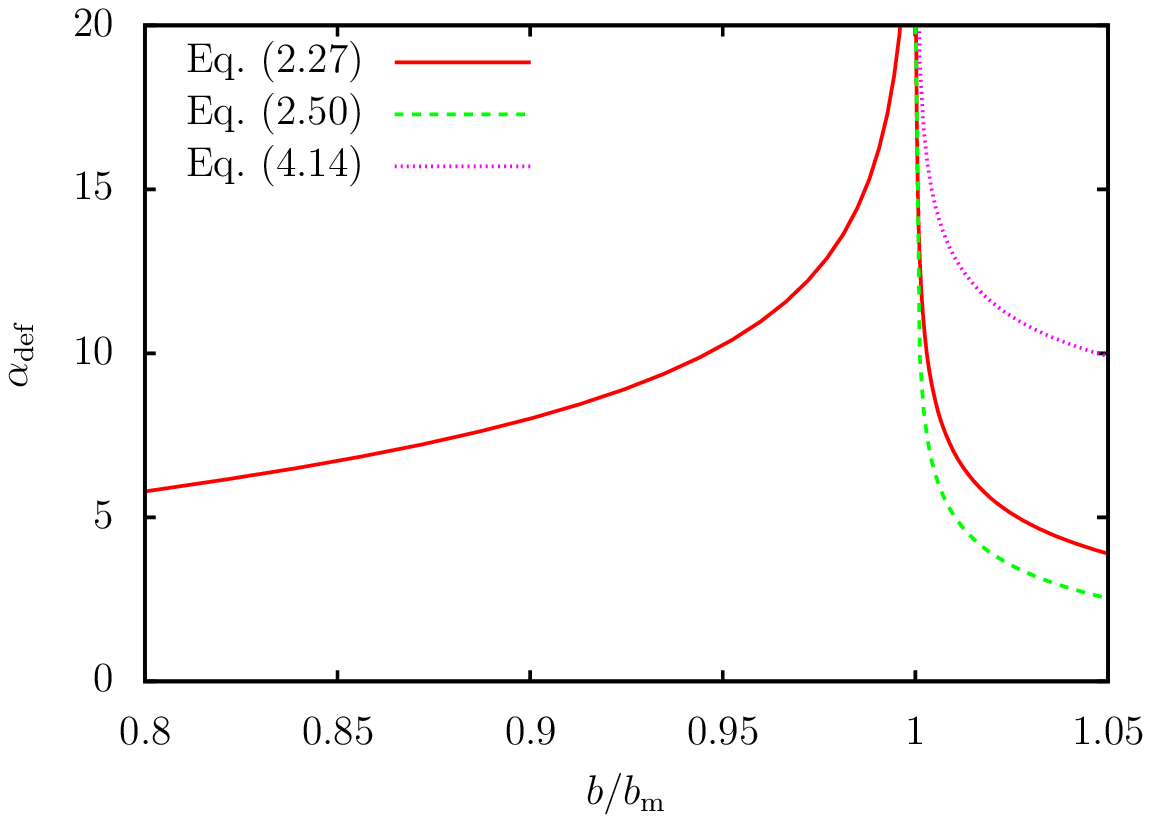}
\end{center}
\caption{The deflection angle as a function of $b/b_{\mathrm{m}}$ in the Hayward spacetime.
The solid~(red), broken~(green), and dotted~(magenta) curves denote the deflection angles~$\alpha_{\mathrm{def}}$ given by Eqs.~(2.27), (2.50), and (4.14), respectively.}
\end{figure}

\section{Discussion and Conclusion}
On this paper, we have obtained the formula of the deflection angle (\ref{eq:def01}) of the light ray, which is reflected by the marginally unstable photon sphere, in the strong deflection limit $b \rightarrow b_{\mathrm{m}}+0$
in a general asymptotically flat, static and spherically symmetric spacetime under some assumptions to calculate observables.
We apply the formula to the Reissner-Nordstr\"{o}m spacetime and the Hayward spacetime with the marginally unstable photon spheres.
The light ray in the strong deflection limit $b \rightarrow b_{\mathrm{m}}+0$ comes from a spatial infinity and it approaches to the marginally unstable photon sphere as $r\rightarrow r_{\mathrm{m}}+0$.
From the effective potentials $V$ in Figs.~2 and 4 in the Reissner-Nordstr\"{o}m and Hayward spacetimes, respectively, 
a light ray in a limit $b\rightarrow b_{\mathrm{m}}-0$, which comes from the spatial infinity and it is reflected by the effective potential, 
also approaches to the marginally unstable photon sphere as $r\rightarrow r_{\mathrm{m}}-0$.
The observables of the light rays in the limit $b\rightarrow b_{\mathrm{m}}-0$ lie outside the scope of this paper 
but results in Ref.~\cite{Shaikh:2019itn} imply that the images with $b\rightarrow b_{\mathrm{m}}-0$ might be brighter than the ones $b\rightarrow b_{\mathrm{m}}+0$ in the winding number $n\geq 1$.
 
We comment on the case of the winding number $n=0$.
In the case, we cannot use the strong deflection limit analysis since its error becomes large. 
Usually we apply the weak-field approximation~$r_0 \ll b$ to solve the lens equation as shown in textbooks~\cite{Schneider_Ehlers_Falco_1992,Schneider_Kochanek_Wambsganss_2006}. 
Note that there is an image with $b \ll b_{\mathrm{m}}$ near the center of the lens object. It can be calculated numerically 
without the strong deflection limit analysis such as the deflection angles calculated numerically in Figs.~3 and 5.

Notice that there is the other pair with a negative impact parameter to each image with a positive impact parameter. 
The magnitudes of the image angles $\theta_{n}$ and magnifications $\mu_n$ of the images with the winding number $n\geq 1$ with the negative impact parameters 
are almost the same as the ones with the positive impact parameters.
The difference of image angles of a couple of images and their total magnification are given by $2\theta_{n}\sim 2\theta_{\mathrm{E}n}$ and $\mu_{n \mathrm{tot}} \sim 2 \left| \mu_{n} \right|$, respectively,
for each $n$ in the strong gravitational fields as shown Table~I.  
If an observer, a lensing object, and a source object are aligned, i.e., $\phi=0$, a couple of images become a (relativistic) Einstein ring. 

We compare the strong deflection limit analysis of the marginally unstable photon sphere with a numerical calculation 
to confirm it.
In Tables~II and III, the image angle $\theta_{\mathrm{SDL}}$ in the strong deflection limit by using Eq.~(2.50) and the image angle $\theta_{\mathrm{NUM}}$ solved numerically by using Eq.~(2.27) 
with the distances $D_{\mathrm{OS}}=16$kpc and $D_{\mathrm{OL}}=D_{\mathrm{LS}}=8$kpc, the mass $M=4\times 10^6 M_{\odot}$, and the source angle $\phi=0$
in the Reissner-Nordstr\"{o}m spacetime and the Hayward spacetime, respectively, are shown for the winding numbers $n=1$, $2$, and $3$.
The errors $(\theta_{\mathrm{NUM}}-\theta_{\mathrm{SDL}})/\theta_{\mathrm{NUM}}$ with winding numbers $n=1,$ $2,$ and $3$ are given by $-10^{-2}$, $-10^{-3}$, and $-10^{-4}$, respectively.
\begin{table*}[htbp]
 \label{table:II}
 \caption{$2\theta_{\mathrm{SDL}}$, $2\theta_{\mathrm{NUM}}$, and $(\theta_{\mathrm{NUM}}-\theta_{\mathrm{SDL}})/\theta_{\mathrm{NUM}}$ for $\theta_{\mathrm{E}1}$, $\theta_{\mathrm{E}2}$, and $\theta_{\mathrm{E}3}$ in the Reissner-Nordstr\"{o}m spacetime. 
 Here, $\theta_{\mathrm{SDL}}$ is the image angle in the strong deflection limit by using Eq.~(2.50) and $\theta_{\mathrm{NUM}}$ is the image angle obtained by using Eq.~(2.27) which is solved numerically.
 The distances $D_{\mathrm{OS}}=16$kpc and $D_{\mathrm{OL}}=D_{\mathrm{LS}}=8$kpc and the mass $M=4\times 10^6 M_{\odot}$ are assumed. 
}
\begin{center}
\begin{tabular}{c c c c} \hline
                                          &$\theta_{\mathrm{E}1}$   &$\theta_{\mathrm{E}2}$  &$\theta_{\mathrm{E}3}$ \\ \hline
 $2\theta_{\mathrm{SDL}}$[$\mu$as]                 &36.8325         &36.5009        &36.4775  \\ 
 $2\theta_{\mathrm{NUM}}$[$\mu$as]                  &37.2827         &36.5504        &36.4871  \\ 
 $\quad(\theta_{\mathrm{NUM}}-\theta_{\mathrm{SDL}})/\theta_{\mathrm{NUM}}\quad $ &-1.21$\times 10^{-2}$         &-1.35$\times 10^{-3}$        &-2.63$\times 10^{-4}$\\ 
 \end{tabular}
\end{center}
\end{table*}
\begin{table*}[htbp]
 \label{table:III}
  \caption{$2\theta_{\mathrm{SDL}}$, $2\theta_{\mathrm{NUM}}$, and $(\theta_{\mathrm{NUM}}-\theta_{\mathrm{SDL}})/\theta_{\mathrm{NUM}}$ for $\theta_{\mathrm{E}1}$, $\theta_{\mathrm{E}2}$, and $\theta_{\mathrm{E}3}$ in the Hayward spacetime.
  The distances $D_{\mathrm{OS}}=16$kpc and $D_{\mathrm{OL}}=D_{\mathrm{LS}}=8$kpc and the mass $M=4\times 10^6 M_{\odot}$ are assumed.}
\begin{center}
\begin{tabular}{c c c c} \hline
                                                      &$\theta_{\mathrm{E}1}$          &$\theta_{\mathrm{E}2}$          &$\theta_{\mathrm{E}3}$ \\ \hline
 $2\theta_{\mathrm{SDL}}$[$\mu$as]                             &46.4819                &46.2617                &46.2461\\ 
 $2\theta_{\mathrm{NUM}}$[$\mu$as]                              &46.8842                &46.3045                &46.2542\\ 
 $\quad(\theta_{\mathrm{NUM}}-\theta_{\mathrm{SDL}})/\theta_{\mathrm{NUM}}\quad $  &-8.58$\times 10^{-3}$  &-9.24$\times 10^{-4}$  &-1.75$\times 10^{-4}$\\ 
 \end{tabular}
\end{center}
\end{table*}

In Ref.~\cite{Tsukamoto:2020uay}, Tsukamoto has considered the deflection angle of a light ray reflected 
by a marginally unstable photon sphere in the strong deflection limit in the Damour-Solodukhin wormhole spacetime~\cite{Damour:2007ap,Nandi:2018mzm,Ovgun:2018fnk,Bhattacharya:2018leh,Ovgun:2018swe} with
\begin{eqnarray}
&&A(r)=1-\frac{2M}{r},\\
&&B(r)=\left[ 1-\frac{3M}{r} \right]^{-1},\\
&&C(r)=r^{2}.
\end{eqnarray}
The deflection angle in the strong deflection limit has a form of Eq.~(\ref{eq:def02}).
See. Eq.~(3.61) in  Ref.~\cite{Tsukamoto:2020uay}.
We notice that the marginally unstable photon sphere $r=r_\mathrm{m}= 3M$ corresponds with a wormhole throat which is satisfied $1/B(r_{\mathrm{m}})=0$
and it violates our assumption that $B(r)$ is positive and finite in a domain $r\geq r_\mathrm{m}$.
Therefore, results on this paper are compatible with ones in Ref.~\cite{Tsukamoto:2020uay}.

In Ref.~\cite{Paul:2020ufc}, Paul has investigated the deflection angle of a light ray in a spacetime, which is suggested by Joshi \textit{et al.} \cite{Joshi:2020tlq}, with 
\begin{eqnarray}
&&A(r)=\frac{1}{B(r)}=\frac{1}{\left( 1+ \frac{M}{r} \right)^2},\\
&&C(r)=r^{2}.
\end{eqnarray}
The spacetime has a naked singularity at $r=0$.
Light rays reach into the curvature singularity 
if $b<b_\mathrm{cr}$, where $b_\mathrm{cr} \equiv M$ is a critical impact parameter, 
while they are scattered if $b>b_\mathrm{cr}$.
The deflection angle in a limit $b \rightarrow b_\mathrm{cr}$ or $r_0 \rightarrow 0$ has a form of Eq.~(\ref{eq:def03}).
See Eq.~(15) in Ref.~\cite{Paul:2020ufc}. 
Since the spacetime does not have a photon sphere at $r=0$ and $A(0)=1/B(0)=0$ there
and since the spacetime does not have a marginally unstable photon sphere,
the spacetime is not satisfied our assumptions.
Thus, results on this paper are compatible with ones in Ref.~\cite{Paul:2020ufc}.

%
\appendix
\section{Expansions in the power of $z$}
The expansions of a function $J(r)$ and its inverse $1/J(r)$ in the power of $z$ are given by
\begin{eqnarray}
J&=&J_0 +J^{\prime}_0 r_0 z+ \left( J^{\prime}_0 r_0 +\frac{1}{2} J^{\prime \prime}_0 r_0^2  \right) z^2 \nonumber\\
&&+ \left( J^{\prime}r_0+ J^{\prime \prime}_0 r_0^2 +\frac{1}{6}J^{\prime \prime \prime}_0 r_0^3 \right) z^3 +O\left( z^4 \right).
\end{eqnarray}
\begin{eqnarray}
\frac{1}{J}&=&\frac{1}{J_0} -\frac{J^{\prime}_0 r_0}{J_0^2}z\nonumber\\
&&+\left( -\frac{J^{\prime}_0 r_0}{J_0^2} +\frac{J_0^{\prime 2} r_0^2}{J_0^3}- \frac{J^{\prime \prime}_0 r_0^2}{2J_0^2} \right) z^2 \nonumber\\
&&+\left( -\frac{J^{\prime}_0 r_0}{J_0^2} +\frac{2J_0^{\prime 2} r_0^2}{J_0^3} -\frac{J_0^{\prime 3} r_0^3}{J_0^4} -\frac{J^{\prime \prime}_0 r_0^2}{J_0^2} \right. \nonumber\\
&&\left. +\frac{J^{\prime}_0 J^{\prime \prime}_0 r_0^3}{J_0^3} -\frac{J^{\prime \prime \prime}_0 r_0^3}{6 J_0^2} \right)z^3 +O\left(z^4 \right).
\end{eqnarray}

\end{document}